\documentclass[reprint,prb]{revtex4-1}

\usepackage{graphicx}
\usepackage{color}
\usepackage{dcolumn}
\usepackage{bm}
\usepackage{amsmath}

\begin{document}
\title{Vortex-density waves and negative absolute resistance in patterned superconductors}

\author{Rog\'erio M. da Silva}
%\email{rogerio@df.ufpe.br}
\affiliation{Departamento de F\'isica, Universidade Federal de Pernambuco, Cidade Universit\'aria, 50670-901 Recife-PE, Brazil}
\author{Cl\'ecio C. de Souza Silva}
%\email{clecio@df.ufpe.br}
\affiliation{Departamento de F\'isica, Universidade Federal de Pernambuco, Cidade Universit\'aria, 50670-901 Recife-PE, Brazil}

\date{\today}

\begin{abstract}

We study theoretically dynamical phases of vortices in superconducting films with arrays of obstacles. By performing a series of molecular dynamics simulations and analytical calculations, we demonstrate the existence of a phase of soliton-like vortex-density waves existing in a wide range of parameters. These waves are formed by a self-assembled phase separation process induced by strongly nonlinear density fluctuations of the moving vortex matter above a certain critical driving current. At high vortex concentrations, the waves move at an approximately current-independent speed resulting in a wide plateau in the voltage-current characteristics. At stronger drives, the vortex system enters into a fully jammed (zero-voltage) phase. By combining ac and dc drives, the interplay between the vortex-density-wave and jammed phases leads to the observation of negative absolute mobility of vortices, which induces the superconducting film into a negative resistance state. 

\end{abstract}


\maketitle

\section{Introduction}

Driven many-particle systems are often affected by strongly nonlinear fluctuations which can lead to dramatic phenomena such as self-organized density waves \cite{Baxter89,Jaeger92}, jamming \cite{HelbingRMP}, and spontaneous segregation \cite{Jaeger92,Poschel95}. In inhomogeneous type-II superconductors, nonlinear phenomena resulting from the motion of magnetic flux quanta (vortices) have been investigated, for instance, in the context of dendritic flux instabilities stemming from the breakdown of the Bean critical state \cite{Leiderer92,Duran95,Colauto10}. 

In a superconductor, vortices are subjected to the Lorentz force induced by an applied current density $J$ and to the pinning force produced by sample inhomogeneities. A dissipative dynamical state of moving vortices is established at the critical current $J_c$ when the Lorentz force overcomes pinning.\cite{Blatter} Just above $J_c$, motion is usually plastic with vortices moving at different velocities, which results in a nonlinear velocity-force (voltage-current) characteristic. At a stronger drive, the pinning potential is washed out and the corresponding dynamical state, coined \emph{flux flow}, is essentially linear. These dynamical regimes have been extensively investigated in disordered superconductors \cite{Blatter,Yeh84} and, more recently, in films with periodic arrays of nanoengineered pinning centers.\cite{Reichhardt97,Kes98,Gutierrez09} However, much less attention has been given to the dynamics of vortices in arrays of obstacles (or antipinning centers).\cite{Golib08} Because in this configuration vortices are not trapped individually, it renders a very small critical current, which, in principle, has limited interest for applications. On the other hand a small $J_c$ allows for investigation of vortex motion in a much wider current range. Moreover, vortices may be forced into meandering paths inducing strong lateral fluctuations that can give rise to novel dynamical phases. 

In this paper, we demonstrate the breakdown of the flux-flow state in a superconducting film with an array of obstacles into a vortex-density wave state and, subsequently, into a fully jammed phase. 
Evidence of density waves in vortex systems have previously been found in the context of current-depaired vortices and antivortices in clean superconductors \cite{Hebboul99} and turbulence in superfluids \cite{Mongiovi07}. %In those systems, the waves are produced by local unbalance of creation and annihilation processes. 
In contrast, the phenomena we present here are collective states of the moving vortex matter resulting from the interplay of vortex-vortex interactions and the friction induced by the obstacles. As we shall see in this paper, these competing interactions lead to a critical profile of the density waves where distinct dynamical phases, as well as a jammed phase, coexist.

The paper is organized as follows. In Sec~\ref{sec.model}, we give the details of our model and numerical procedure. In Sec.~\ref{sec.DynPh}, we present the main results of our molecular dynamics simulations and discuss the dynamical phases. An analytical model is proposed in Sec.~\ref{sec.MF} to explain the conditions for formation and stabilization of vortex-density waves and how this leads to a constant voltage regime. Sec.~\ref{sec.MF} is devoted to illustrating an application of the interplay between the moving and jammed phases, namely, the negative absolute resistance effect. Finally, our main findings and final remarks are summarized in Sec.~\ref{sec.concl}.

\section{Model and numerical details} \label{sec.model}

We consider a set of $N$ vortices generated in the film by a perpendicular magnetic field $\vec{B}=B\hat{z}$. The dynamics of a vortex $i$ is modeled by the Bardeen-Stephen equation,  
\begin{equation}\label{eq.motion}
 \eta \vec{v}_i = \vec{F} -\vec{\nabla}_iU_b - \sum_{j\neq i}\vec{\nabla}_iU_{ij}\,,
\end{equation}
where $\eta$ is the viscous drag coefficient, $\vec{F}$ is the Lorentz force induced by the applied current, and $U_{ij}$ is the vortex pair potential, modeled here as $U_{ij}=\epsilon K_0(r_{ij}/\Lambda)$ ($\epsilon = \phi_0^2/(4\pi\mu_0\Lambda)$ and $\phi_0$ is the flux quantum). $\Lambda$ is the effective penetration depth measuring the range of vortex-vortex interactions. For $\Lambda$ much larger than all length scales of the system $U_{ij}$ asymptotically reduces to a logarithmic potential. In this investigation, we analyze the effect of short and long range interactions by conducting simulations for $0.5\leq\Lambda\leq\infty$. Hereafter we adopt the following units: $a$ (obstacle lattice constant) for length, $t_0=\epsilon/(\eta a^2)$ for time, and $\epsilon$ for energy.  

\begin{figure}[t]
 \includegraphics[width=0.65\columnwidth]{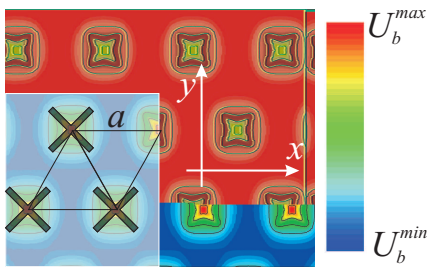}
 \caption{(Color online) Foreground: unit cell of the triangular  array of ferromagnetic barriers. Background: contour plot of the potential $U_b(x,y)$ generated by the array. Each cross comprises two legs perpendicular to one another, of length $0.5a$, width $0.04a$, and height $0.025a$. To prevent proximity effects, the ferromagnets are assumed to be separated from the superconductor by a $0.0125a$ thick insulating layer.}
 \label{fig.Pot}
\end{figure}

$U_b(x,y)$ is the potential resulting from a triangular array of cross-shaped obstacles (Fig.~\ref{fig.Pot}). 
To be specific we have chosen as such obstacles small ferromagnets (FMs), each with a uniform permanent magnetization $\vec{M}$ antiparallel to $\vec{B}$. This choice is motivated by the well-known properties of FMs to either attract or repel vortices, depending on their magnetic orientation~\cite{Morgan98}. Other possible choices include higher-$T_c$ superconducting dots~\cite{Gillijns07} and pillars~\cite{Golib08}. Each FM repels a vortex $j$ via a potential given by $U_{kj}=-\int \vec{M}\cdot\vec{b}_jd^3r_k$, where $\vec{r}_k=(x_k,y_k)$ is the position of the volume element of the micromagnet $k$ and $b_j$ is the local flux density of vortex $j$ at $\vec{r}_k$ calculated within the London approximation~\cite{Milosevic02,Carneiro05}. In the results presented here we used $M=320\phi_0/\mu_0a^2$ ($=0.8B_{c2}/\mu_0$, assuming a coherence length $\xi=0.02$). For this value the ferromagnets do not induce any vortex-antivortex pairs~\cite{Lima09}. $U_b(x,y)$ results from the superposition of all FMs in the array (Fig.~\ref{fig.Pot}). 
It introduces a current-induced trapping mechanism between two critical drive values: $F_{c1}=3.5$, at which a vortex can get trapped at a cross corner, and $F_{c2}=12.0$, above which a vortex surmounts the barrier. These properties provide the main ingredient for the phenomena we investigate here, namely, \emph{a non-linear, drive-dependent friction}.

Eqs. \ref{eq.motion} are solved numerically via molecular dynamics (MD) simulations in a cell of size $L_x\times L_y$ with periodic boundary conditions. The values of $L_x$ and $L_y$ were chosen after a careful finite size analysis. We verified that the transitions between the observed dynamical phases become unaltered for system sizes typically larger or equal to 24$\times$48 or 12$\times$96 FMs. Therefore, $N$ typically ranges up to a few thousands of vortices. The simulation procedure is as follows. First, the vortex system is equilibrated at zero current via a standard simulated annealing scheme~\cite{Lima09,Clecio06}. Then, the driving force $F_y$, applied along the $y$ axis is slowly increased. For each $F_y$, time series and averages are calculated on an interval $\Delta t$ of typically $10^6$ time steps after a stationary state has settled. The main physical quantities are the center-of-mass velocity $v_y = \frac{1}{N}\sum_i\vec{v}_i\cdot\hat{y}$ of the vortex array along the drive direction and its time average.

\section{Dynamical phases} \label{sec.DynPh}

\subsection{General picture} \label{sec.DynPh-general}

\begin{figure}[b]
\includegraphics[width=0.9\columnwidth]{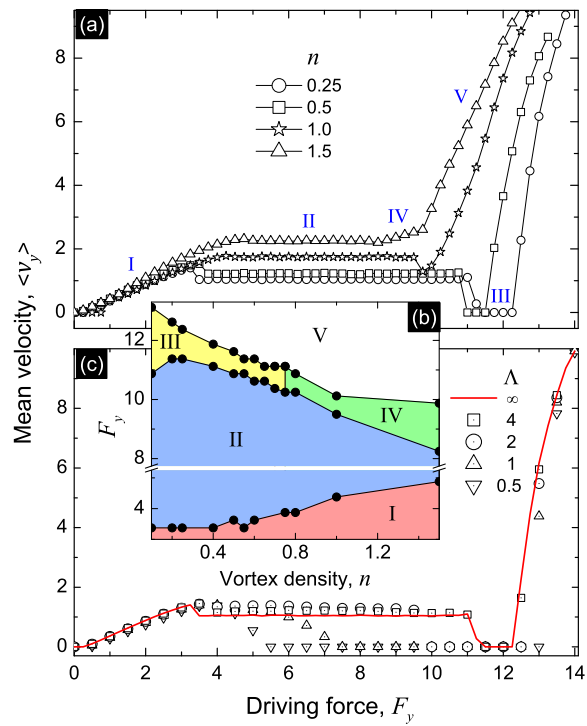}
\caption{(Color online) (a) Vortex mean velocity $\langle v_y\rangle$ as a function of the driving force $F_y$ for logarithmic ($\Lambda=\infty$) vortex-vortex interaction and occupation numbers ranging from $n=0.25$ to 1.5. The roman numbers correspond to the phases depicted in the $F_y$-$n$ diagram (b) and explained in the text. (c) $\langle v_y\rangle$ versus $F_y$ for $n=0.25$ and different $\Lambda$ values: $\Lambda=0.5$ to 4 (symbols) and $\Lambda=\infty$ (line).}
\label{fig.VI}
\end{figure}

We conducted an extensive series of simulations of the model described above for densities ranging from $n=0.1$ to 2.0 vortices per obstacle and different values of $\Lambda$. Fig.~\ref{fig.VI}-(a)  
presents $\langle v_y\rangle$-$F_y$ curves for a few $n$ values. The different dynamical behaviors are arranged in the diagram of Fig.~\ref{fig.VI}-(b). In general, for $F\lesssim F_{c1}$, all vortices meander through the potential channels leading to a linear (flux-flow) phase, which we call phase I. This phase has an effective viscous drag coefficient, defined as $\eta_{\rm eff} = F/\langle v \rangle$, approximately force independent and smaller than $\eta$. This is consistent with the fact that the meandering motion of vortices dissipates more energy as compared to the case of conventional flux flow. 

\begin{figure*}[ht]
\includegraphics[width=1\textwidth]{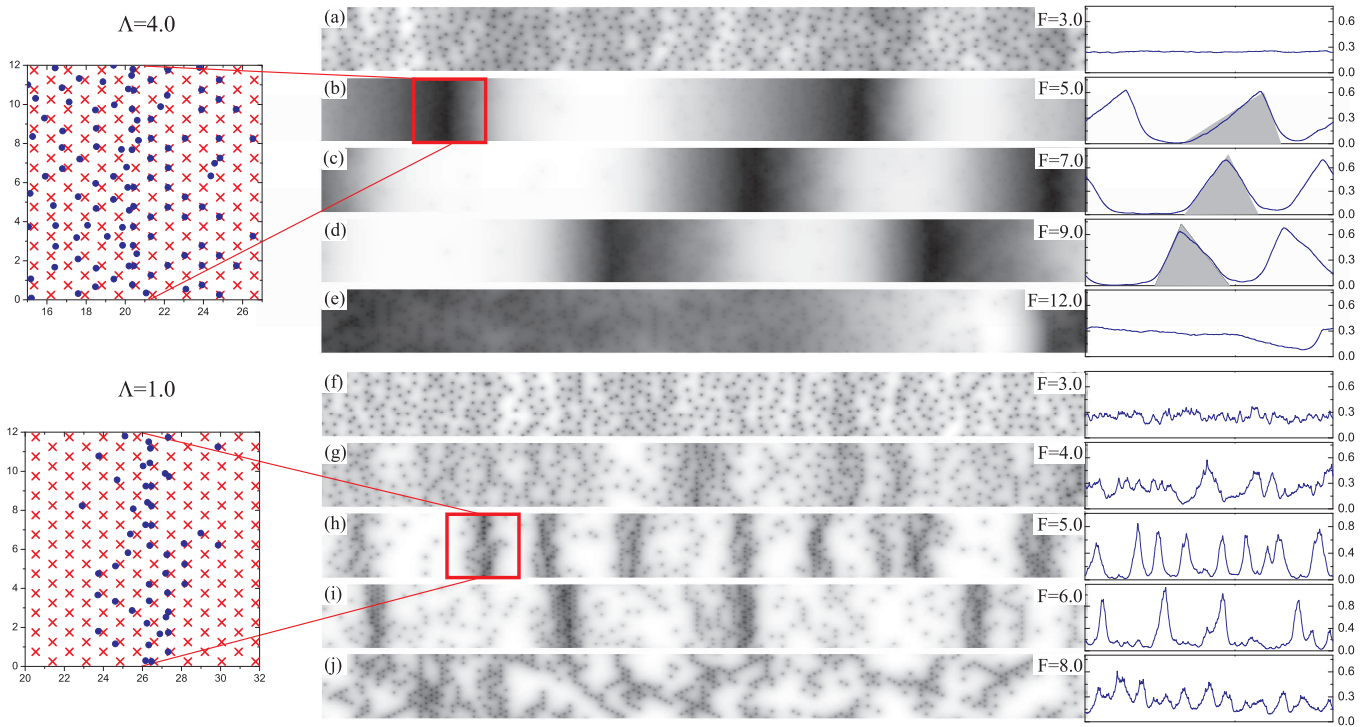}
\caption{(Color online) Snapshots of the flux density distribution $b(x,y)$ in a 12$\times$144 array of cross barriers for total vortex density $n=0.25$, $\Lambda=4.0$ [(a)-(e)] and 1.0 [(f)-(j)], and different force values spanning phase I [(a) and (f)], II [(b)-(d) and (g)-(i)], and III [(e) and (j)]. The drive direction ($y$ axis) points left to right. The curves on the right depict the flux density $b(y)$ (in units of $\phi_0$ per unit cell) integrated over the sample width. The shaded areas represent the shape of the density wave as predicted by the mean-field model presented in Sec.~\ref{sec.MF}. The plots on the left represent a zoom-in of the regions indicated by the squares showing the positions of vortices (dots) and barriers (crosses). }
\label{fig.flux}
\end{figure*}

Above $F_{c1}$, we observed a remarkable plateau in $\langle v_y\rangle(F_y)$ for long-ranged vortex-vortex interactions. This constant velocity state, phase II, persists in a wide range of force values and ends drastically in a fully jammed ($\langle v_y\rangle$=0) state for $n<0.75$, phase III, or in a moving linear phase IV for larger $n$. At even higher drives ($F\gtrsim F_{c2}$), all vortices surmount the barriers leading to a new flux-flow regime (phase V). Phase II dominates the diagram in the range of parameters studied. To check the stability of this phase with respect to the kind of vortex-vortex interaction, we fixed $n=0.25$ and run simulations for $\Lambda$ values down to 0.5. 
%$0.5\leq\Lambda<\infty$
As observed in Fig.~\ref{fig.VI}-(c), the plateau shrinks as $\Lambda$ decreases and is substituted by a smooth decrease of $\langle v_y\rangle$ for $\Lambda\leq 1.0$. As we shall see below, this region of decreasing $\langle v_y\rangle$ have similar properties to the plateau region of the $\Lambda\geq 2$ systems.

\subsection{Vortex-density waves} \label{sec.DynPh-VDW}

Here we analyze phase II in more detail. For simplicity we focus only on the vortex density $n=0.25$. In Fig.~\ref{fig.flux}, we plot snapshots of the flux density distribution $b(x,y)$ for $\Lambda=4$ (long range) and $\Lambda=1$ (short range). The central panels correspond to density plots of $b(x,y)$ whereas the panels on the right are flux profiles averaged over the $x$ direction, i.e., $\bar{b}(y)=\int\!dx\,b(x,y)/L_x$. The left panels depict the snapshots of vortex positions in the areas indicated. These plots reveal a spontaneous breakdown of the vortex distribution when going from phase I [panels (a) and (f)] to phase II [(b)-(d), for $\Lambda=4.0$, and (g)-(i), for $\Lambda=1.0$]. The resulting patterns comprise stripes of high flux density oriented perpendicularly to the drive direction and traveling at constant speed as soliton-like vortex-density waves (VDW). 

For long range interactions, the waves have a peculiar triangular flux profile, which changes shape as the force is increased, and are intercalated by narrow regions of sparsely distributed vortices trapped by the cross corners. By carefully analyzing the dynamics in different regions of the sample, we found out that the left side of the wave is characterized by meandering motion of vortices, similar to the dynamics at $F<F_{c1}$ (phase I), while on the right side vortices will rather assist each other to jump over the barrier, which is the same dynamics found at $F>F_{c2}$ (phase V). Therefore, a single density wave corresponds to a moving region where two distinct dynamical phases coexist. Such phase separation was also observed for short range interactions, but not so clearly because in this case a density wave spans only a few lattice spacings of the FM array. In addition, the pinned vortex regions for small $\Lambda$ are considerably larger and increases with force, which is accompanied by a reduction of the mean vortex speed. A detailed analysis of the morphology of the VDWs and its dependence on the driving force and other parameters is given in Sec.~\ref{sec.MF}. 

Although the vortex wave patterns observed in our simulations are not space periodic, they do present time order. A convenient way to search for time correlations in the vortex density $n(t)$ is by studying the power spectrum of its fluctuations, $S(f)=\langle|n(f)|^2\rangle$, with $n(f)=\int\!dt\,n(t)e^{-i2{\pi}ft}$.~\cite{Yeh84} We perform this task by recording the density $n(t)$ integrated over a $12\times 12$ region of the film every other 10 time steps during a series of $10^6$ time steps. $S(f)$ is then estimated by averaging $|n(f)|^2$ over $N_s=10$ segments of the time series. 

\begin{figure}[t]
\includegraphics[width=0.85\columnwidth]{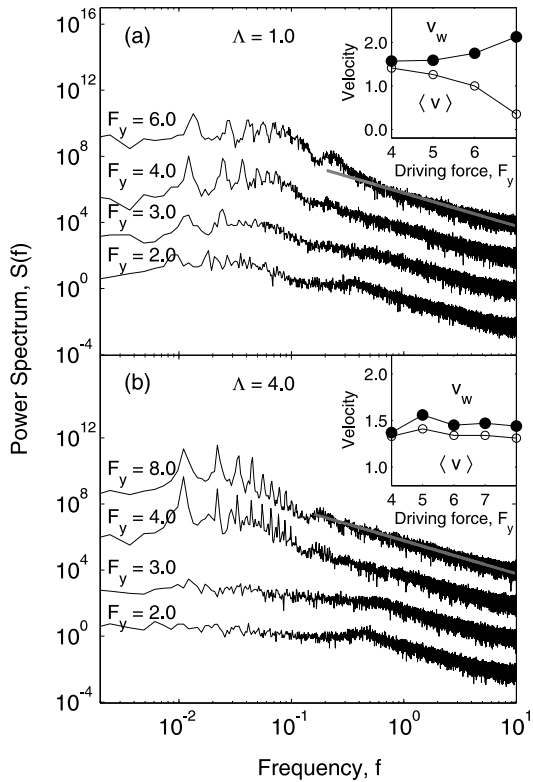}
\caption{Power spectra of density fluctuations at different drives for $n=0.25$ and $\Lambda=1.0$ (a) and $\Lambda=4.0$ (b). The curves are displaced from each other by two decades for better visualization. In both panels $1/f^2$ curves (gray line) are plotted as guides to the eye (see text). Insets: driving force dependence of the mean vortex velocity and the density wave velocity for $\Lambda=1.0$ and 4.0.}
\label{fig.power}
\end{figure}

Plots of typical $S(f)$, for both short ($\Lambda=1.0$) and long ($\Lambda=4.0$) range cases, are depicted in Fig.~\ref{fig.power}. 
As it is clear in this figure, $S(f)$ has a long $1/f^2$ tail for all drive intensities and interaction ranges studied. This high frequency behavior is associated with Brownian noise resulting mainly from the meandering  dynamics, where a vortex colliding with a barrier has to choose randomly whether to dodge left or right. The low frequency region of $S(f)$, however, changes dramatically from a flat behavior in phase I to a series of strong peaks in phase II. These peaks correspond to the harmonics associated with the coherent passage of high flux density domains at constant speed. Therefore, the power spectra provide an efficient means of determining the density-wave velocity: $v_w=f_1L_y$, where $f_1$ is the first harmonic frequency. 

We have calculated the wave speed for different force values both for $\Lambda=1.0$ and $4.0$. For better accuracy, we calculate $v_w$ from a (sharper) higher harmonic $f_s=sf_1$ (typically, $s=3$ or 4). 
The results are plotted in the insets of Fig.~\ref{fig.power}. Remarkably, $v_w$ is always larger than the mean vortex speed for both kinds of interactions. For short-range interaction, $v_w$ increases with the driving force, whereas $\langle v\rangle$ decreases. Such high speeds of the density waves as compared to $\langle v\rangle$ can be explained by the presence of slow moving vortices between adjacent waves. In fact, because vortices in a density valley are diluted, they can easily be trapped at barrier corners (thus contributing to decrease $\langle v\rangle$) until being catch by a wave front. In opposition, vortices in the density peaks assist each other to overcome the traps. Put in another way, the VDWs observed here are \emph{anti-jamming waves} self-assembled in a way as to avoid a full stop of the vortex flow. 

Density waves have been observed in other classical many-particle systems, for instance, car traffic~\cite{Kerner93,Kerner98}, granular flow~\cite{Moriyama98,Ellingsen2010} and galaxies~\cite{Tassev08}, and their origin is still unclear. In contrast to the anti-jamming waves observed in our simulations, the density peaks in these systems are jamming domains intercalated by free motion regions. Accordingly, their wave velocities are usually smaller than the mean flow velocity and can even become negative~\cite{HelbingRMP}. However, in all these systems, ours included, density waves can be understood as a dynamical, self-organized phase separation, similar to equilibrium pattern formation in systems with competing interactions~\cite{Seul95}. For instance, in the case of granular media flowing through narrow tubes~\cite{Ellingsen2010} density waves form spontaneously and generate a spectrum of density fluctuations very similar to those presented here in Fig.~\ref{fig.power}, in spite of the very different types of interaction in these systems and different nature of the waves. 

\section{Mean-field approach} \label{sec.MF}

To understand how the vortex-density waves are formed and why they lead to a constant voltage regime for large $\Lambda$, we propose an analytical model based on a mean-field analysis of Eq.~\ref{eq.motion}. We start from the Fokker-Planck equation for the vortex distribution function at zero temperature and coarse-grain it over scales larger than $\Lambda$. This procedure leads to the following equation for the vortex density $\rho$:\cite{Zapperi2001}
\begin{equation}\label{eq.F-P}
 \eta\frac{\partial\rho}{\partial t} = -\vec{\nabla} \cdot \left[ 
 \rho\left( \vec{F} - g\vec{\nabla}\rho + \vec{F}_P \right) \right].
\end{equation}
$- g\vec{\nabla}\rho$ is the lowest order (long-wavelength) contribution of vortex-vortex interactions, with $g=-\frac{1}{2}\int d^2r\,\vec{r}\cdot \vec{\nabla}U_{ij}$. This indicates that, as expected, a local flux gradient introduces an additional term to the local driving force, that is $\vec{F}_d=\vec{F} - g\vec{\nabla}\rho$. $\vec{F}_P$ is the coarse-grained force produced by the pattern, which can be understood as a drive-dependent friction force. 

To account for a dynamical trapping mechanism similar to that induced by the cross-shaped barriers, we assume that $F_P$ represents a static friction in the range $F_{c1}<F_d<F_{c2}$, for which no motion occurs. For $F_d\leq F_{c1}$ and $F_d\geq F_{c2}$, motion takes place and, accordingly, $F_P$ must represent the corresponding kinetic frictions ${F}_{k1}$ and ${F}_{k2}$, respectively. These properties can be arranged by modeling the friction force as $\vec{F}_P=-\hat{y}F_P(F_d)$, with:
\begin{equation}\label{eq.friction}
    {F}_P=
    \begin{cases}
      {F}_{k1}, & \text{if}\ F_d\leq F_{c1} \\
      {F}_{k2}, & \text{if}\ F_d\geq F_{c2} \\
      {F}_d, & \text{if}\ F_{c1}<F_d<F_{c2}.
    \end{cases}
\end{equation}

We proceed by assuming a one-dimensional density wave is formed at a given $F$, with $F_{c1}<F<F_{c2}$, and search for the conditions for such a wave to exist with a fixed shape and a constant velocity $\vec{v}_w$ directed parallel to $\vec{F}=\hat{y}F$. In this case, we can transform into a reference frame moving with the wave, $(y'=y-v_wt,t'=t)$, and rewrite Eq.~\ref{eq.F-P} as 
\begin{equation}\label{eq.vw}
 \eta{v}_w = {F} - g\partial_{y'}\rho - F_P,
\end{equation}
where $\rho$ is assumed to be a function of $y'$ only and  $\partial_{y'}\rho$ is the density gradient in the moving frame. To determine the shape of the wave, it is convenient to choose the position of the wave maximum as the origin of $y'$ and analyze separately the regions $l=\{y'<0\}$ and $r=\{y'>0\}$. Further, for simplicity, we assume the wave is single-peaked and define $G_{r(l)}\equiv|\partial_{y'}\rho|_{r(l)}$, such that $F_d = F-gG_l$, for $y'<0$, and $F_d = F_L+gG_r$, for $y'>0$. 

With the help of Eq.~\ref{eq.friction}, we can state the following conditions of motion of the wave:
\begin{align}
  gG_l & \geq F-F_{c1}, \label{eq.ineq1}\\
  gG_r & \geq F_{c2}-F, \label{eq.ineq2}
\end{align}
where we used the fact that, for any $F\in[F_{c1},F_{c2}]$, the positive slope on the right side of the wave rules out the possibility $F_d>F_{c2}$, whereas the negative slope on the right rules out $F_d<F_{c1}$. This naturally accounts for the dynamical phase separation observed in our MD simulations and discussed in Sec.~\ref{sec.DynPh-VDW}. 

A direct consequence of the above condition is that if the vortex distribution peaks up at a certain point, the corresponding bump can only turn into a traveling wave if the gradients in both sides of the bump fulfill Eqs.~\ref{eq.ineq1} and \ref{eq.ineq2}. Subcritical bumps, i.e. those with $gG_l < F-F_{c1}$ and $gG_r < F_{c2}-F$, would simply be trapped by the pattern. On the other hand, an increase in $G_r$ above its critical value leads to an increase in the right side velocity while an increase in $G_l$ above $F_{c2}-F$ reduces velocity on the left side. Therefore, if the vortex distribution peaks up too sharply the wave will be unstable and will rapidly decay to a profile with gentler slopes so that the velocities on the left and right sides of the wave come to a common value. It is clear then that the condition for the stability of a wave traveling at constant speed and fixed shape is that \emph{the density gradient must be everywhere critical}, that is
\begin{align}
	G_l & = \left(F_L-F_{c1}\right)/g, \label{eq.critL} \\
	G_r & = \left(F_{c2}-F_L\right)/g.\label{eq.critR}
\end{align}
This result is analogous to the build up of the Bean critical state in the magnetization of hard type-II superconductors, where the flux gradient (induced by the external flux penetrating the material) exactly balances the critical force induced by pinning centers. Here the critical forces are exactly balanced by the overall driving force, which accounts for the applied Lorentz force and the flux gradient.

Eqs.~\ref{eq.critL} and \ref{eq.critR} describe a triangular shape for the density wave which switches its left (right) side slope from gentle (steep) to steep (gentle) as the Lorentz force is increased from $F_{c1}$ towards $F_{c2}$. To contrast this result with our simulation data, we take $F_{c1}=3.5$ and $F_{c2}=12$, obtained from the $\langle v \rangle(F)$ characteristic for $\Lambda=4$ and $n=0.25$, and calculate the parameter $g$ for Bessel-like vortex-vortex interactions, which gives $g=2\pi\Lambda^2\epsilon$. Then, we calculate the critical gradients from Eqs.~\ref{eq.critL} and \ref{eq.critR} and compare with the flux profiles obtained from the MD simulations for $F=5.0$, 7.0, and 9.0. The results, presented in Fig.~\ref{fig.flux} (b)-(d), demonstrate a good agreement between the numerics and the mean-field model. 

To estimate the wave velocity, we first impose that the velocity in the left side of the wave ($v_l=v_w$) must be consistent with the linear, meandering dynamics expected to occur for $F_d<F_{c1}$. For this kind of motion, an increase in the drive leads to a proportional increase in the velocity, i.e., $v_w = F_d/\eta_\text{eff}$. Then, we notice that, from Eq.~\ref{eq.critL}, $F_d=F-gG_l=F_{c1}$. Hence, the wave velocity does not depend on $F$: 
\begin{equation}
	v_w=F_{c1}/\eta_\text{eff}. \label{eq.vw2}
\end{equation}
This is in excellent agreement with the MD results for large $\Lambda$ if one assumes that all vortices participate on a wave, that is $\langle v \rangle=v_w$. In fact, as revealed by the simulations, there are narrow, low-density regions of trapped vortices coexisting with the dynamical phases for $n=0.25$ and large $\Lambda$. For that reason, $v_w$ is typically slightly larger than $\langle v \rangle$. 

It is worth noticing that in the limit $F_{c1}\rightarrow 0$ a moving wave solution is not possible (at least in the framework of the present model). Therefore, a condition for the occurrence of vortex-density waves is that vortices interact with a potential which is able to trap them only when the applied current surpass a certain critical value. This is probably the reason why vortex-density waves have never been detected or predicted in, for instance, arrays of pinning centers. In these systems, in general, there is no moving phase preceding the pinned phase.

It is also important to emphasize that the mean-field approximation (Eq.~\ref{eq.F-P}) is strictly valid for vortex densities $n\gg\Lambda^{-2}$. In the simulations performed for $\Lambda=4a$ and $n=0.25\frac{\sqrt{3}}{2}a^{-2}$ we have $n=3.46\Lambda^{-2}$, which can be considered inside the range of validity of the model. Indeed, the main results obtained in the MD calculations for these parameters were correctly reproduced by the mean-field approach. For $\Lambda=a$, however, the vortex distribution is too diluted. In addition, in order to treat the interaction with the barriers macroscopically, $\Lambda$ should span several lattice spacings. For these reasons, the analytical results presented here are not strictly valid for short range interactions. Notwithstanding, our simple analysis ignores the details of the barriers, taking into account only its main macroscopic properties. Therefore, it can be applied to other barrier configurations, as long as their coarse-grained properties can be expressed in a way similar to Eq.~\ref{eq.friction}.

\section{Negative absolute resistance}

The dynamical transition between the VDW and jamming phases opens the possibility for an interesting application: the construction of an active device (made of superconducting material) exhibiting \emph{negative absolute resistance} (NAR). The working principle of this device is based on the negative absolute mobility (NAM) effect, which corresponds to motion in a direction that opposes the driving force, irrespective of the drive direction. This phenomenon was predicted to occur in the transport of a single classical Brownian particle through a symmetric, periodic substrate~\cite{Eichhorn2002} and was experimentally corroborated in a system of colloidal spheres in a microfluidic device~\cite{Ros2005} and subsequently in a Josephson junction~\cite{Nagel08}. In the later, the phase dynamics, which can be mapped into the problem of a single Brownian particle, lead to negative resistance. 

Quite generally, the main ingredients for the observation of NAM are: (i) a medium that allows easy motion of the particles at low drives and strongly suppresses mobility at high drives; (ii) a fluctuating (nonequilibrium) force, for instance an ac excitation, superposed to the dc drive. The first requirement implies necessarily that, for a certain force range, the mean velocity of the driven system must decrease as the driving force increases, that is, negative differential mobility. This phenomenon have been experimentally demonstrated for vortices driven in superconducting films with periodic arrays of pinning centers at magnetic fields close to one flux quantum per pinning site ($B_1$).\cite{Gutierrez09} Previous numerical calculations have pointed out that such negative differential mobility of vortices near $B_1$ is a result of a transition from a highly-dissipative, disordered regime to a filamentary state with somewhat smaller but non-vanishing mean velocities.\cite{Reichhardt97,Misko} As shown in Fig.~\ref{fig.VI}, the $VI$ characteristics of the vortex system under investigation here do present negative differential resistance, in a broad field range, as a result of the transition from the VDW to the jammed state. Here, however, the voltage drop corresponds to a strong suppression of vortex mobility, which, as discussed below, is a crucial feature for the observation of NAM.

\begin{figure}[t]
\includegraphics[width=0.80\columnwidth]{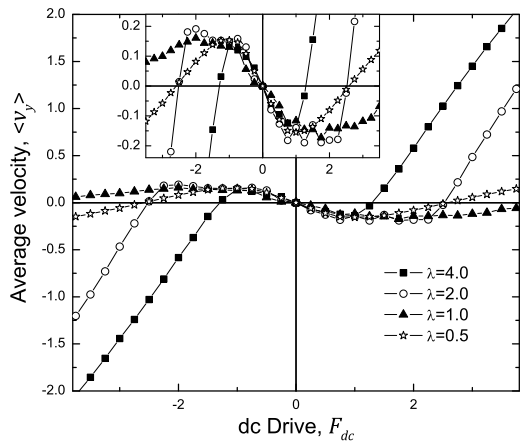}
\caption{Vortex mean velocity as a function of the dc force $F_{dc}$ in the presence of a sinusoidal excitation with frequency $\omega=0.01$ and amplitudes $A=11.25$,$10.25$, $7.5$ and $5.5$ corresponding to $\lambda = 4, 2, 1, 0.5$, respectivelly. All curves exhibit absolute negative mobility of vortices at small $F_{dc}$.}
\label{fig.NAR}
\end{figure}

Here we demonstrate how the NAM effect can be applied to Abrikosov vortices in patterned superconductors in order to achieve negative resistance. We subject the vortices to a Lorentz force $F_y = F_{dc} + F_{ac}(t)$, with $F_{ac}(t)=A\sin(\omega t)$ and an ac amplitude $A$ fixed at a value chosen just above the VDW-jamming (II-III) transition, and measure $\langle v_y\rangle$, which is proportional to the dc voltage, as a function of the dc drive $F_{dc}$, proportional to the dc current. The ac term of $F_y$ provide the extra energy source required for the observation of NAM. The results are shown in Fig.~\ref{fig.NAR} 
for several $\Lambda$ values. At zero dc current ($F_{dc}=0$) the vortex dynamics cycle symmetrically through phases I, II, III, and back, producing zero dc voltage. However, for small positive (negative) $F_{dc}$, vortices are most of the time in the pinned phase III during the positive (negative) half-cycle of $F_{ac}$ and always in one of the moving phases I or II in the negative (positive) half-cycle. This leads to a net motion contrary to the dc force, that is, the mean electric field induced by vortex motion points antiparallel to the applied dc current, thus generating the negative absolute resistance effect.

\section{Conclusions} \label{sec.concl}

In conclusion, we studied novel dynamical phases of vortices in a patterned superconducting film. By molecular dynamics calculations we demonstrated, for a wide range of parameters, the existence of vortex-density waves propagating at a constant speed $v_w$ larger than the mean vortex velocity $\langle v \rangle$ and essentially force-independent for long-range vortex-vortex interactions. The waves consist of well-defined regions of coexisting moving phases, where vortices assist each other to either dodge or overcome the barriers, and are intercalated by regions of trapped vortices. 

Our mean-field analysis revealed that these waves stem from fluctuations in the vortex distribution induced by a highly nonlinear friction force. A density bump produced by such fluctuations turns into a stable moving wave when it reaches a certain critical profile. It is precisely this critical shape of the waves that results in a force independent wave velocity, thereby naturally accounting for the voltage plateau observed in the MD simulations. 

We have also demonstrated the feasibility of a superconducting device which exhibits negative absolute resistance. This is achieved by exploring a combination of ac and dc excitations in such a way as to conveniently switch the system between the moving and fully-jammed phases. This prediction could be promptly tested using conventional transport measurement techniques on a nanostructured sample with a pattern similar to that proposed here.

Finally, it is worth pointing out that, given the generality of our mean field model, the main results predicted here can also be applied to other systems of interacting particles, such as colloids and pedestrians. An advantage of the vortex system proposed here is that interactions can easily be tuned. In typical nanostructured superconducting films, $\Lambda$ can be varied from a fraction to several lattice spacings by controlling the film temperature near $T_c$~\cite{Grigorenko03,Clecio06}, thereby allowing for experimentally accessing our predictions in both long and short range cases. In additional modern imaging techniques could be used to identify the vortex density waves.

\acknowledgements

We would like to thank Alejandro V. Silhanek,  Giovani L. Vasconcelos, and Leonardo R. E. Cabral for useful suggestions and enlightening discussions. This work was supported by the Brazilian science agencies CNPq and FACEPE (grant no. APQ-0589-1.05/08).

%%%%%%%%%%%%%%%% Reference List %%%%%%%%%%%%%%%%%%%%

\end{document}